\documentclass[prl,twocolumn,superscriptaddress,float,aps]{revtex4-2}
\usepackage{graphicx}
\usepackage{epstopdf}
\usepackage{color}
\usepackage{hyperref}
\usepackage{bm}
\usepackage{printlen}
\usepackage{units}
\usepackage{bbm,pifont}
\usepackage{tikz}
\usepackage{amsmath}
\usepackage{amssymb}
\usepackage{amsthm}
\usepackage{commath}
\usepackage{amsfonts}
\usepackage[utf8]{inputenc}
\usepackage{bbold}
\usepackage{dcolumn}
\usepackage{epsfig}
\usepackage{latexsym}
\usepackage{xcolor}
\usepackage{subfigure}
\usepackage{array}
\usepackage{bbm}
\usepackage{hyperref,hypcap}
\usepackage{cancel}
\usepackage{ulem,dsfont}
\usepackage{braket}

\newcommand{\<}{\langle}

\renewcommand{\>}{\rangle}

\renewcommand{\v}[1]{\mathbf{#1}} 

\renewcommand{\d}{\partial}

\begin{document}

\title{Nonreciprocal phonons in $\mathcal{P}\mathcal{T}$-symmetric antiferromagnet}

\date{\today}
\author{Yafei Ren}
\affiliation{Department of Physics and Astronomy, University of Delaware, Newark, Delaware 19716, USA}

\author{Daniyar Saparov} 
\affiliation{Department of Physics, The University of Texas at Austin, Austin, Texas 78712, USA}

\author{Qian Niu} 
\affiliation{ICQD and School of Physics, University of Science and Technology of China, Hefei, Anhui 230026, China}

\begin{abstract}
Phonon nonreciprocity, indicating different transport properties along opposite directions, has been observed in experiments under a magnetic field.
We show that nonreciprocal acoustic phonons can also exist without a magnetic field nor net magnetization. We focus on $\mathcal{P}\mathcal{T}$ symmetric antiferromagnets that break both time-reversal $\mathcal{T}$ and inversion symmetry $\mathcal{P}$. We identify crucial contributions in phenomenological elastic theory, dubbed flexo-viscosity and flexo-torque, that induce phonon nonreciprocity without changing the phonon polarization. The microscopic origin of these contributions is the molecular Berry curvature, manifested as emergent nonlocal magnetic fields on phonons.
The symmetry breaking originated from spin order is transferred to the phonon system through spin-orbit coupling, where the orbital degree of freedom affects the lattice dynamics directly. By electrically modifying the spin-orbit coupling, we show that both the phonon nonreciprocity and helicity can be controlled and enhanced.
Importantly, the phonon nonreciprocity is an odd function of the N\'eel vector, serving as an indicator of the order parameter. 
\end{abstract}

\maketitle

Nonreciprocity, the directional transport of particles or waves~\cite{tokura2018nonreciprocal}, is pivotal for both fundamental research and technological applications. One prime example is the optical nonreciprocity~\cite{sounas2017non, asadchy2020tutorial, huang2021loss, szaller2013symmetry}, which plays an essential role in optical communication and information processing by enabling devices like directional amplifiers~\cite{malz2018quantum}, optical isolators~\cite{jalas2013and} and circulators~\cite{kamal2011noiseless}. 
The nonreciprocity also extends to other bosonic systems, like magnons~\cite{sato2019nonreciprocal, hayami2022essential, gitgeatpong2017nonreciprocal, matsumoto2020nonreciprocal, el2023antiferromagnetic, guckelhorn2023observation} and phonons~\cite{nomura2019phonon,xu2020nonreciprocal,sengupta2020phonon, atzori2021magneto}. Notably, the phonon nonreciprocity facilitates the rectification of heat and sound, vital for advancements in phononics~\cite{sato2019nonreciprocal, li2012colloquium, xu2020nonreciprocal, maldovan2013sound, fornieri2017towards, nassar2020nonreciprocity}.
Besides artificial structures~\cite{nassar2020nonreciprocity} and surface acoustic waves~\cite{xu2020nonreciprocal, shao2022electrical}, nonreciprocal phonons have been studied in magnetic materials with chiral structures or semimetals under a finite magnetic field~\cite{nomura2019phonon, nomura2023nonreciprocal, atzori2021magneto}. 
However, exploring nonreciprocity in the absence of a magnetic field and net magnetization, particularly controllable via an electric field, offers greater scalability and application versatility~\cite{shao2022electrical}.

In this Letter, we study nonreciprocal phonons in $\mathcal{P}\mathcal{T}$ symmetric antiferromagnets (AFMs)~\cite{ahn2024progress} where both inversion ($\mathcal{P}$) and time-reversal ($\mathcal{T}$) symmetries are independently broken, enabling nonreciprocal phenomena. 
We developed a phenomenological elastic theory for nonreciprocal phonons using symmetry principle, where we identify crucial contributions, dubbed flexo-viscosity and flexo-torque, that lead to stress proportional to the gradient of strain rate and rotation rate. Both contributions are odd under $\mathcal{P}$ or $\mathcal{T}$ while even under $\mathcal{P}\mathcal{T}$. Symmetry breaking originates from spin order, which is transferred to phonons through molecular Berry curvature in a microscopic theory~\cite{mead1979determination, qin2012berry, hu2021phonon, saparov2022lattice, bonini2023frequency, ren2024adiabatic}. Spin-orbit coupling plays a crucial role in generating this curvature. By applying an electric field to modify it, the phonon nonreciprocity can be controlled and much enhanced. 
In addition, an electric field breaks the joint $\mathcal{P}\mathcal{T}$ symmetry, enabling phonon helicity and new contributions to the nonreciprocity from Hall viscosity and strain gradient.
% in a microscopic mechanism that allows phonons to display \textit{inherent} nonreciprocity

To highlight the difference from previously reported nonreciprocal phonons where phonons exhibit \textit{inherited} nonreciprocity~\cite{gitgeatpong2017nonreciprocal, matsumoto2020nonreciprocal, cheong2018broken} via coupling to nonreciprocal magnons~\cite{hayami2016asymmetric, gitgeatpong2017nonreciprocal, matsumoto2020nonreciprocal, hayami2022essential}, we neglected the spin dynamics or the magnons and show that phonons can display \textit{inherent} nonreciprocity through the molecular Berry curvature.
Moreover, the phonon nonreciprocity switches as the N\'eel vector flips. The close relationship between phonon nonreciprocity and magnetic orders enables the interplay of phononics, spintronics, and magnonics in $\mathcal{P}\mathcal{T}$ symmetric AFMs. 

\textit{\textbf{Symmetry analysis.---}} We examine two-dimensional materials with three-fold rotation symmetry~\cite{ahn2024progress} as the nonreciprocity requires the breaking of even-fold rotation symmetry about $z$ axis. Phonon nonreciprocity is characterized by the difference between the velocities at momentum $\bm{k}$ and $-\bm{k}$, $\delta \bm{v}(\bm{k})$, which is a periodic function of the direction of $\bm{k}$. 
Consequently, $\delta \bm{v}(\bm{k})$ vanishes along specific directions, which is set as the $y$-direction and its counterparts by rotation of $\pm 2\pi/3$.

%This can also be a result of the joint symmetry $\mathcal{M}_{x} \mathcal{T}$ that combines the mirror reflection $\mathcal{M}_{x}$ and time-reversal operation.

The long-wavelength acoustic phonons can be described by the elastic theory, with the leading terms captured by the following Lagrangian
\begin{align}
    \mathcal{L}_0 =& \frac{1}{2} \rho \dot{\bm{u}}^2 - 2\mu\left[\frac{1}{4}(\epsilon_{xx}-\epsilon_{yy})^2 + \epsilon_{xy}^2\right] \notag \\
    &- 2(\mu+\lambda) (\epsilon_{xx}+\epsilon_{yy})^2 
\end{align}
where $\rho$ represents the mass density, $\dot{\bm{u}}$ the time-derivative of the displacement field $\bm{u}$, and $\epsilon_{ij}=\frac{1}{2}(\partial_i u_j + \partial_j u_i)$, with $\mu$ and $\lambda$ as Lam\'e factors. 

The breaking of $\mathcal{P}$ and $\mathcal{T}$ symmetries manifests in higher-order corrections (see detailed derivations in the Supplemental Material~\cite{footnote_unit}):
\begin{align}
    & \tau^H \frac{1}{2}(\dot{\epsilon}_{xx}+\dot{\epsilon}_{yy}) \left[ \frac{1}{2}\partial_x ({\epsilon}_{xx}-{\epsilon}_{yy})-\partial_y \epsilon_{xy} \right] \\
   +& \tau^M \dot{m}_{xy} \left[ \frac{1}{2}\partial_y ({\epsilon}_{xx}-{\epsilon}_{yy})+\partial_x \epsilon_{xy} \right].
   \nonumber
\end{align}
Here $m_{ij}=\frac{1}{2}(\partial_i u_j - \partial_j u_i)$ represents the rotation tensor. The term $\tau^H$ introduces a stress related to the gradient of the strain rate, leading to a force depending on the lattice velocity, and is thus termed ``flexo-viscosity''. 
This stress is dissipationless, as the flexo-viscosity tensor is antisymmetric~\cite{barkeshli2012dissipationless}. 
The term $\tau^M$ denotes a ``flexo-torque'', a force depending on the gradient of the rotation rate $\dot{m}_{ij}$, indicating a rotation-induced stress or a strain-gradient driven lattice rotation. 
Both terms break $\mathcal{P}$ and $\mathcal{T}$ symmetries, fostering phonon nonreciprocity, yet remain invariant under the joint $\mathcal{P}\mathcal{T}$ operation. 
This ensures that the phonon modes do not carry angular momentum showing linear polarization, as phonon helicity is $\mathcal{P}\mathcal{T}$ odd.

When external electric or magnetic fields break the joint $\mathcal{P}\mathcal{T}$ symmetry, they enable new contributions to the phonon nonreciprocity. On the one hand, a higher-order elastic energy, which is $\mathcal{P}$-odd but $\mathcal{T}$-even, can be nonzero: $ \zeta \left[ -\epsilon_{xx} \partial_y \epsilon_{yy}+(\epsilon_{xx}+\epsilon_{yy}) \partial_x \epsilon_{xy} \right] $. This term, which involves stain and its gradient, leads to stress proportional to the strain gradient and is termed ``flexo-elastic energy'', with $\zeta$ as the flexo-elastic modulus. Although the higher-order elastic theory, including the second-order deformation gradient, was well established in the strain-gradient theory~\cite{mindlin1965second, hutchinson1997strain,  lam2003experiments}, flexo-elastic energy is novel as previous study focused on isotropic media. 
On the other hand, they also allow the presence of $\mathcal{P}$-even but $\mathcal{T}$-odd corrections: $\eta^{H} \dot{\epsilon}_{xy}({\epsilon}_{xx}-{\epsilon}_{yy}) + \eta^{M} \dot{m}_{xy}({\epsilon}_{xx}+{\epsilon}_{yy})$.
The term $\eta^H$ corresponds to the phonon Hall viscosity~\cite{barkeshli2012dissipationless}, which modifies the stress tensor by introducing velocity-dependent forces through the strain rate $\dot{\epsilon}_{kl}$, and is also dissipationless~\cite{barkeshli2012dissipationless}. Unlike the Hall viscosity, the $\eta^M$ term indicates that the time-varying of the rotation angle, corresponding to a nonzero phonon angular momentum, leads to an effective bulk stress. In reciprocal, it also indicates that bulk stress can lead to a change of rotation angle, indicating a torque on the mechanical motion. We thus term it ``stricto-torque'' to highlight the coupling between dynamical rotation with the mechanical deformation.
In isotropic media (or with four-fold rotation symmetry around $z$-axis), symmetry guarantees that $\eta^H=\eta^M$~\cite{barkeshli2012dissipationless}. In the anisotropic case considered here, $\eta^H$ and $\eta^M$ are independent. 

Together, flexo-elastic energy and corrections like the Hall viscosity or stricto-torque contribute to breaking both the inversion and time-reversal symmetries. 
This leads to additional phonon nonreciprocity components proportional to $\zeta (\eta^H+\eta^M)$ as shown later. Moreover, breaking $\mathcal{P}\mathcal{T}$ symmetry enables a nonzero phonon angular momentum, causing phonon modes to exhibit elliptical polarization due to the hybridization between the longitudinal and transverse modes. 

\textit{\textbf{Nonreciprocal phonon modes.---}} Here, we examine the influence of the previously discussed terms on the phonon nonreciprocity and normal modes. We focus on the dispersion along $\hat{x}$ direction with $\bm{k}=k\hat{x}$ and express the normal mode as $\bm{u} = \bm{u}_{\bm{k}}e^{-i\omega t + i \bm{k}\cdot \bm{r}}$ where $\bm{u}_{\bm{k}}=(u_{\bm{k},x}, u_{\bm{k},y})^T$ represents the polarization vector. The equation of motion becomes $\frac{\rho}{2}\omega^2 \bm{u}_{\bm{k}} = (\mathcal{D}_0 + \delta \mathcal{D}) \bm{u}_{\bm{k}}$
where $\mathcal{D}_0$ is the dynamical matrix from the first-order elastic theory described by $\mathcal{L}_0$ and $\delta \mathcal{D}$ is the higher-order correction. Here
\begin{align}
    \mathcal{D}_0=& \left[
    \begin{array}{cc}
    (\mu + \frac{\lambda}{2})k^2  & 0 \\
    0 & \frac{\mu}{2} k^2 
    \end{array}
    \right] 
\end{align}
generates the conventional longitudinal and transverse modes with linear polarization and linear dispersion, $\omega_{0,\alpha}=v_{\alpha} |k|$, where $\alpha=\rm{L}$ or $\rm{T}$ for longitudinal and transverse modes, respectively. Phonon velocities are $v_{\rm L}=\sqrt{(2\mu+\lambda)/\rho}$ and $v_{\rm T}=\sqrt{\mu/\rho}$. The higher-order correction is
\begin{align}
    \delta \mathcal{D}=& \left[
    \begin{array}{cc}
    \frac{\tau^H}{4} \omega k^3    & - \frac{i\zeta}{4}k^3 - i\frac{\eta}{4}\omega k^2 \\
    \frac{i\zeta}{4}k^3 + i\frac{\eta}{4}\omega k^2     &  \frac{\tau^M}{4} \omega k^3
    \end{array}
    \right]
\end{align}
where $\eta=\eta^H+\eta^M$.

In $\mathcal{P}\mathcal{T}$-symmetric antiferromagnets, both $\zeta$ and $\eta$ vanish. Nonzero $\tau^{H,M}$ modifies the phonon dispersion directly without altering the phonon normal modes. The longitudinal phonon dispersion becomes $\omega_{\rm L}=\omega_{0,\rm L}+\frac{\tau^H}{4\rho}k^3$, ignoring higher order corrections to $k^5$. The phonon nonreciprocity is characterized by the velocity difference, $\delta v_k=|v(k)|-|v(-k)|$, quantified as $\delta v_k/v_{\rm L} = 3\tau^H k^2/2\rho v_{\rm L}$. Similarly, the dispersion of the transverse mode is also nonreciprocal: $\delta v_k/v_{\rm T} = 3\tau^M k^2/2\rho v_{\rm T}$. It is noted that the flexo-Hall viscosity and flexo-torque do not change the phonon polarization, consistent with the $\mathcal{P}\mathcal{T}$-symmetry. 

In the absence of this symmetry, both $\zeta$ and $\eta$ are nonzero. The phonon dispersion up to cubic order in $k$ is 
\begin{align}
    \omega_{\rm L} = \omega_{0,{\rm L}} +&  \frac{\zeta^2+(\eta v_{\rm L})^2}{8\rho^2 v_{\rm L} (v_{\rm L}^2-v_{\rm T}^2)} |k|^3 \nonumber \\
    +& \frac{\zeta\eta}{4\rho^2 (v_{\rm L}^2-v_{\rm T}^2)} k^3 +\frac{\tau^H}{4\rho}k^3 \\
    \omega_{\rm T} = \omega_{0,{\rm T}} -&  \frac{\zeta^2+(\eta v_{\rm T})^2}{8\rho^2 v_{\rm T} (v_{\rm L}^2-v_{\rm T}^2)} |k|^3 \nonumber \\
    -& \frac{\zeta\eta}{4\rho^2 (v_{\rm L}^2-v_{\rm T}^2)} k^3 +\frac{\tau^M}{4\rho}k^3. 
\end{align}
One can find that when both $\zeta$ and $\eta$ are nonzero, an additional contribution to the phonon nonreciprocity appears, proportional to $\zeta\eta$. 
The corrections proportional to $|k|^3$ from either flexo-elastic energy or the Hall viscosity are reciprocal. 
In addition, unlike the $\mathcal{P}\mathcal{T}$-symmetric case, nonzero $\zeta$ or $\eta$ couples the longitudinal and transverse modes, leading to elliptical phonon polarization and nonzero angular momentum.

\textit{\textbf{Microscopic origin.---}} Molecular Berry curvature offers a microscopic physical mechanism, by which the symmetry breaking of the electronic ground state, induced by the magnetic order, transfers to the lattice dynamics~\cite{mead1979determination, qin2012berry, hu2021phonon, saparov2022lattice, bonini2023frequency, ren2024adiabatic}.
This curvature stems from the adiabatic evolution of the electronic ground state following the lattice vibration where the spin-orbit coupling is crucial~\cite{saparov2022lattice}. The resulting geometrical Berry phase acts on the lattice dynamics as a gauge field analogy to a magnetic field. Such feedback effect can be described by the following effective Lagrangian of the lattice dynamics
\begin{align}\label{LatticeLag}
    \mathcal{L} = \sum_{l,\kappa} \frac{M_{\kappa}}{2}\dot{\bm{u}}_{l,\kappa}^2  + \bm{A}_{l,\kappa}\cdot\dot{\bm{u}}_{l,\kappa}  - V_{\rm eff}(\{\bm{u}\})
\end{align}
where the first term is the kinetic energy with $\bm{u}_{l,\kappa}$ being the displacement of the sublattice $\kappa$ in the $l$-th unit cell and $M_{\kappa}$ being the atomic mass. 
$V_{\rm eff}$ is the scalar spring potential from the electronic energy whereas $\bm{A}_{l,\kappa}$ is a vector potential, as it couples to the lattice velocity $\dot{\bm{u}}_{l,\kappa}$, that arises from the electronic ground state wavefunction $|\Psi_0\rangle$ with $\bm{A}_{l,\kappa}=i\langle \Psi_0 |\partial_{{\bm{u}}_{l,\kappa}}\Psi_0 \rangle$.
The associated molecular Berry curvature is given by~\cite{saparov2022lattice}
\begin{align}
    G^{l\kappa \alpha}_{l'\kappa' \alpha'} = 2 \text{Im}\Big\<\frac{\d \Phi_0}{\d u_{l', \kappa'\beta}} \Big| \frac {\d \Phi_0}{\d u_{l, \kappa\alpha}} \Big \>
\end{align}
where $|\Phi_0\rangle$ is the electronic ground state that depends on the atomic displacement $u_{l,\kappa\alpha}$ along the $\alpha$-direction. The $|\Phi_0\rangle$ carries the symmetry-breaking information.
The equation of motion reads
\begin{align}
    M_\kappa \ddot{u}_{l,\kappa\alpha} = - \frac{\partial V_{\rm eff}}{\partial u_{l,\kappa\alpha}} - \sum_{l'\kappa' \alpha'} G^{l\kappa \alpha}_{l'\kappa' \alpha'} \dot{u}_{l'\kappa' \alpha'}.
\end{align}
A nonzero molecular Berry curvature breaks the time-reversal symmetry of the equation. The velocity-dependent force is an analogy of the Lorentz force and is dissipationless as the $G^{l\kappa \alpha}_{l'\kappa' \alpha'}$ matrix is anti-symmetric by exchanging the super and subscript. Moreover, in contrast to the Lorentz force, this force is nonlocal as the force on atom $\{l\kappa\}$ depends on the velocity of other atoms.
By employing the Fourier transform and a generalized Bogoliubov transformation, one can obtain the phonon dispersion and polarization~\cite{saparov2022lattice}.

\begin{figure}
\includegraphics[width=8 cm]{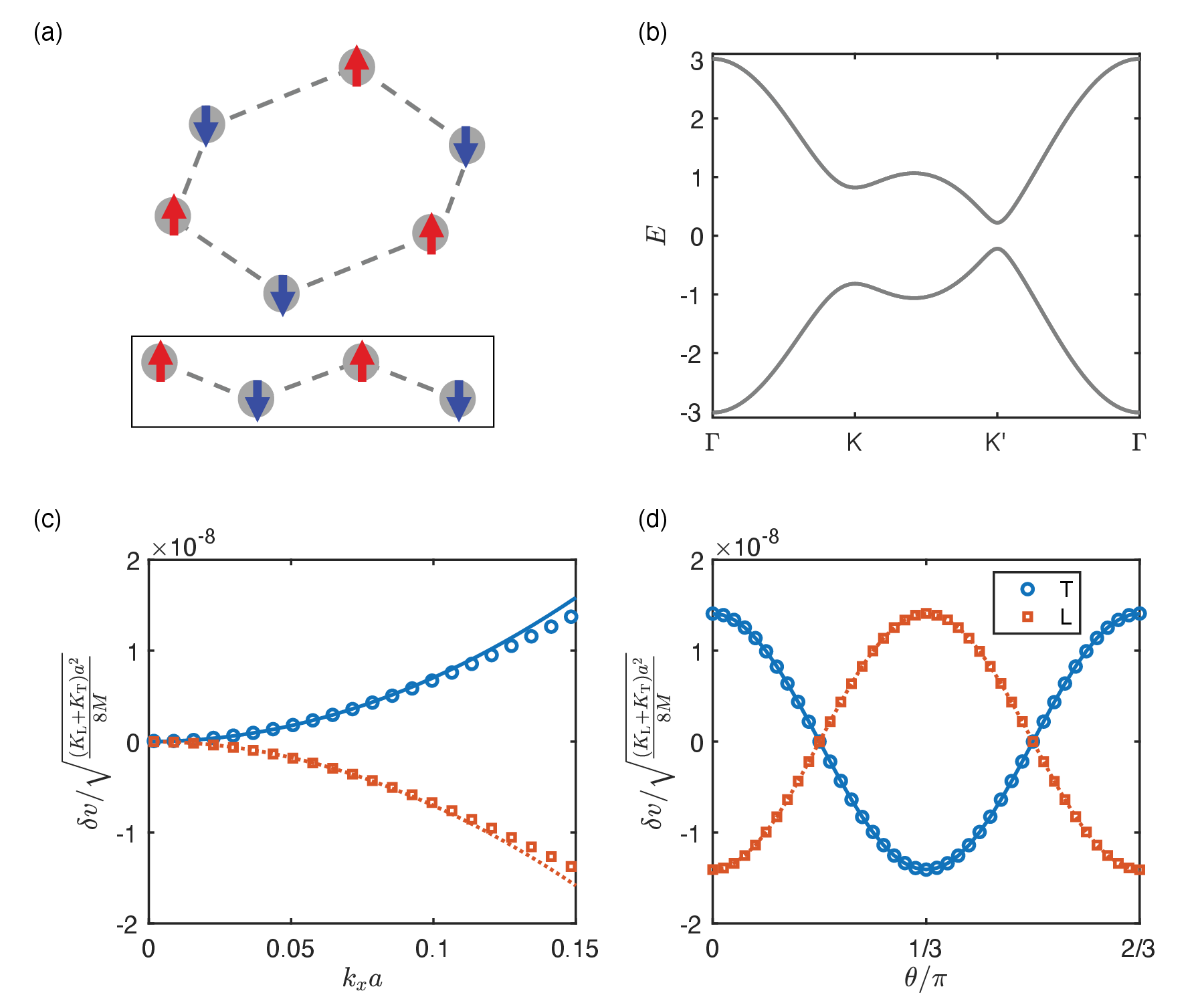} 
\caption{(a) Lattice structure of a $\mathcal{PT}$-symmetric antiferromagnet on a buckled honeycomb lattice. (b) Electronic band structure. (c) Phonon nonreciprocity $\delta v=|v(\bm{k})|-|v(-\bm{k})|$ vs momentum amplitude. Here $\bm{k}=(k_x,0)$. The transverse and longitudinal branches are shown in circle and square, separately. Lines are guides of the eye obtained by quadratic fitting. (d) Phonon nonreciprocity $\delta v$ vs angle $\theta$. Here $\bm{k}=k_0(\cos\theta,\sin\theta)$. Lines are guides of the eye that are proportional to $\cos 3\theta$.}
\label{PhononAFM}
\end{figure}

\textit{\textbf{A case study.---}} Here we demonstrate the phonon nonreciprocity in a $\mathcal{P}\mathcal{T}$ symmetries AFM. We consider a buckled honeycomb lattice that hosts an easy-axis N\'eel order as illustrated in Fig.~\ref{PhononAFM}(a). The localized $d$ electrons responsible for the magnetic order interact with valence electrons through $s$-$d$ exchange interaction. The corresponding Hamiltonian is given by:
\begin{align}
H =&  -t \sum_{\langle ij \rangle} c_i^{\dagger}c_j + i \lambda_{\rm I}\sum_{\langle\langle ij \rangle \rangle} \nu_{ij}  c_i^{\dagger} s_{z} c_j \\ 
 -& i\lambda_{\rm IR} \sum_{\langle\langle ij \rangle\rangle } \mu_{ij} c_i^\dagger (\bm{s}\times \hat{\bm{d}}_{ij})_z c_j 
 + \lambda_{\rm N} \sum_i \mu_{ii} c_i^{\dagger}s_{z}c_i \nonumber
\end{align}
where $c_i^\dagger=(c^\dagger_{i,\uparrow}, c^\dagger_{i,\downarrow})$ are creation operators of spin up and down electrons on the site $i$. $-t$ represents the nearest neighbor hopping energy, $\lambda_{\rm I}$ denotes the strength of the Kane-Mele type intrinsic spin-orbit coupling with $\nu_{ij} = {\rm sign}\left[ (\hat{\bm{d}}_{1} \times \hat{\bm{d}}_{2})_{z} \right] = \pm1 $, $ \hat{\bm{d}}_{1}$ and $\hat{\bm{d}}_{2}$ as the unit vectors of the two bonds connecting site $i$ to $j$. $\bm{s}=(s_x, s_y, s_z)$ are spin Pauli matrices. The third term stands for the intrinsic Rashba spin-orbit coupling from the buckling of the honeycomb lattice that breaks the two-fold rotation symmetry about the $z$-axis~\cite{liu2011low} with $\mu_{ij}=\pm 1$ for $\{i,j\}\in $ A or B sublattice, respectively.
The fourth term represents the staggered exchange field that arises from the N{\'e}el order with an out-of-plane easy axis. This term breaks both $\mathcal{T}$ and $\mathcal{P}$ symmetry while respecting the joint symmetry of $\mathcal{P}\mathcal{T}$. This symmetry guarantees the double degeneracy of the energy bands as shown in Fig.~\ref{PhononAFM}(b). 

Spin-orbit coupling plays a pivotal role, linking spin to the electrons’ orbital motion, which in turn interacts directly with lattice dynamics through the dependence of the hopping integrals on the lattice displacement (see more details in Supplemental Material~\cite{footnote_unit}).

The lattice dynamics is described by a mass-spring model. We assume the masses of A and B sublattices are the same, denoted as $M$. The nearest neighboring sites are connected by springs with spring constant along the bond direction (longitudinal) being $K_L$ and the spring constant perpendicular to the bond direction (transverse) being $K_T$. 

Numerical calculation confirms the nonreciprocity in this $\mathcal{P}\mathcal{T}$-symmetric system. Figure~\ref{PhononAFM}(c) shows $\delta v_k=|v(\bm{k})|-|v(-\bm{k})|$ as a function of momentum $\bm{k}=(k,0)$. One can find a quadratic velocity difference at small $k$, consistent with the theoretical analysis. The opposite sign of $\delta v_k$ for longitudinal and transverse modes indicates opposite signs of $\tau^H$ and $\tau^M$. Figure~\ref{PhononAFM}(d) shows $\delta v_k=|v(\bm{k})|-|v(-\bm{k})|$ as a function of the orientation of $\bm{k}=k_0 (\cos \theta, \sin \theta)$ at $k_0=0.3/a$ with $a$ being the lattice constant. It shows a three-fold rotation symmetry with maximized amplitude along the $x$ direction and zero value along the $y$ direction.

\begin{figure}
\includegraphics[width=8 cm]{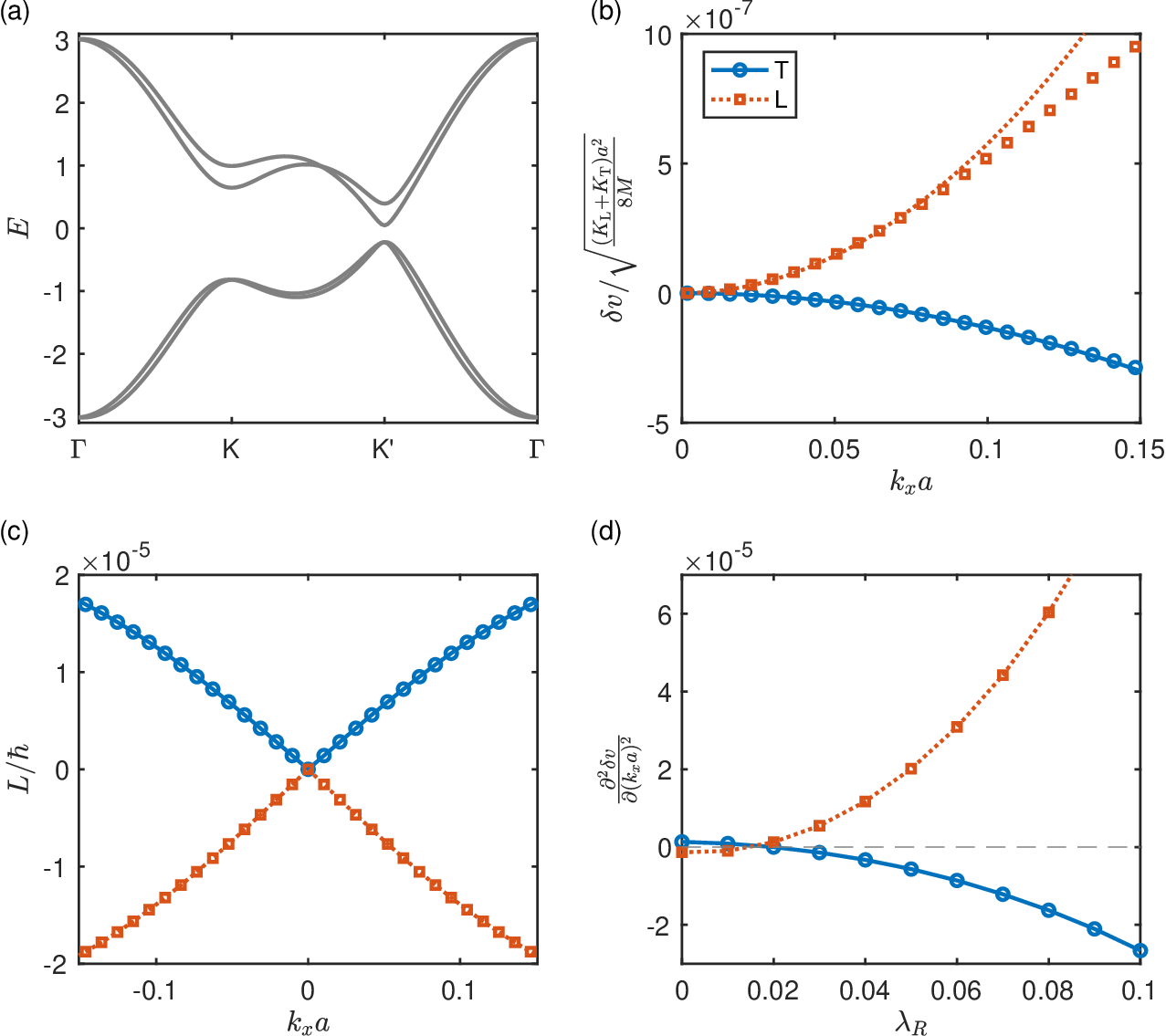} 
\caption{(a) Electronic band structure with Rashba spin-orbit coupling. (b) Velocity difference $\delta v$ vs momentum $k_x a$. (c) Phonon angular momentum $L$ vs momentum $k_x a$. (d) Dependence of $\partial^2 \delta v/\partial (k_x a)^2$ on the strength of Rashba spin-orbit coupling.}
\label{PhononAFM1}
\end{figure}

\textit{\textbf{Electric tunability.---}} 
The phonon nonreciprocity can be controlled by breaking the $\mathcal{P}\mathcal{T}$-symmetry using external means, like an electric field, a magnetic field, or a substrate. Specifically, an out-of-plane electric field induces a Rashba spin-orbit coupling
\begin{equation}
H' = i \lambda_{\rm R} \sum_{\langle ij \rangle} c_i^{\dagger} (\bm{s} \times \hat{\bm{d}}_{ij})_{z} c_j
\end{equation}
where $\lambda_{\rm R}$ represents the spin-orbit coupling strength, proportional to the electric field. 
This term lifts the spin degeneracy of the energy bands as shown in Fig.~\ref{PhononAFM1}(a). 
Phonon nonreciprocity is also altered as shown in Fig.~\ref{PhononAFM1}(b). Unlike the results in Fig.~\ref{PhononAFM}(c), the longitudinal phonon here has a larger velocity along $x$ direction. Additionally, the phonon angular momentum has also changed to nonzero values as demonstrated in Fig.~\ref{PhononAFM1}(c).

By adjusting the amplitude of Rashba spin-orbit coupling, $\lambda_{\rm R}$, the phonon nonreciprocity is calculated systematically. 
Figure~\ref{PhononAFM1}(d) demonstrates how the $d^2 \delta v_k/dk^2$ varies with $\lambda_{\rm R}$. The derivative is calculated at $k=0$ along $x$ direction. One can find that the phonon nonreciprocity shows a strong dependence on the amplitude of $\lambda_{\rm R}$, even exhibits a sign change. However, it is independent of the sign of $\lambda_{\rm R}$, i.e., the direction of the electric field. This is because the systems with opposite electric fields are related to each other by a $\mathcal{PT}$ operation. Under this operation, $\delta v_k$ is invariant. Thus, it is thus an even function of the electric field.
Nevertheless, the phonon angular momentum depends on the sign of the electric field as the phonon angular momentum changes sign under $\mathcal{PT}$ operation. 

Additionally, phonon nonreciprocity and angular momentum are sensitive to the orientation of the N\'eel vector. At a fixed electric field, reversing the N\'eel vector through time-reversal operation alters the signs of both the phonon nonreciprocity and phonon angular momentum. Therefore, they can serve as indicators of the N\'eel vector orientation.

\textit{\textbf{Summary.---}}
We explored the nonreciprocal acoustic phonons in $\mathcal{PT}$-symmetric AFMs, highlighting the absence of magnetic fields. The breaking of both inversion and time-reversal symmetries by spin order, conveyed through spin-orbit coupling, facilitates phonon nonreciprocity. A phenomenological elastic theory identifies two key mechanisms, dubbed flexo-viscosity and flexo-torque, that contributes to the nonreciprocity. The microscopic origin is captured by molecular Berry curvature, which enables inherent phonon nonreciprocity, independent of magnons. Control over phonon nonreciprocity is achievable through an external electric field, which modifies the molecular Berry curvature by influencing electrons orbital motion. Phonon nonreciprocity, being an odd function of the N\'eel vector, serves as a potential indicator of the order parameter.

While we neglect the magnon-phonon coupling here, this coupling has a deep connection with the molecular Berry curvature. In this study, the spin dynamics is neglected by fixing the orientation of the N\'eel vector. When this constraint is relaxed by allowing the spin wavefunction to evolve adiabatically following acoustic lattice vibration, one can obtain additional contributions to the molecular Berry curvature~\cite{bonini2023frequency, ren2024adiabatic}. 
Moreover, the time evolution of the spin wavefunction further generates Berry curvature in the parameter space spanned by spin orientation, which serves as a microscopic origin of the Landau-Lifshitz equation, leading to spin dynamics and thus magnons~\cite{niu1998spin}. 

\begin{acknowledgments}
\textit{Acknowledgements.---}
Y.R. acknowledges the helpful discussion with Prof. Di Xiao and Prof. David Vanderbilt. Y.R. is supported by the startup funds provided by the College of Arts and Sciences and the Department of Physics and Astronomy of the University of Delaware. Q.N. is supported by the National Natural Science Foundation of China (Grant No. 12234017) and the National Key Research and Development Program of China (Grant No. 2023YFA1406300). 
\end{acknowledgments}

\end{document}